\documentclass[prd,twocolumn,showpacs,floatfix,nofootinbib]{revtex4-1}

\usepackage{hyperref}
\usepackage{latexsym}
\usepackage{graphicx,epsf,epsfig}
\usepackage{amssymb,amsmath}
\usepackage{bm}
\usepackage[usenames]{color}
\usepackage{pstricks}
\usepackage{rotating}
\usepackage[utf8]{inputenc}
\usepackage[T1]{fontenc}

\newcommand{\PR}[1]{\ensuremath{\left[#1\right]}}
\newcommand{\PC}[1]{\ensuremath{\left(#1\right)}}
\newcommand{\chaves}[1]{\ensuremath{\left\{#1\right\}}}

\graphicspath{{./figures/}}

\begin{document}

\title{Statistical mechanics of self-gravitating systems: \\ mixing as a criterion for indistinguishability}
\author{Leandro {Beraldo e Silva}$^{1}$}
\email{lberaldo@if.usp.br}
\author{Marcos Lima$^{1}$}
\author{Laerte Sodr\'e$^{2}$}
\author{J\'er\^ome Perez$^{3}$}
\affiliation{
$^{1}$ Departamento de F\'isica Matem\'atica, Instituto de F\'isica, Universidade de S\~ao Paulo, S\~ao Paulo-SP, Brazil \\
$^{2}$ Departamento de Astronomia, Instituto de Astronomia, Geofísica e Ciências
  Atmosféricas, Universidade de São Paulo, S\~ao Paulo-SP, Brazil \\
$^{3}$ Laboratoire de Mathématiques Appliquées, ENSTA Paristech, Paris, France
}
\date{\today}

\begin{abstract}
  We propose an association between the phase-space mixing level of a self-gravitating system and
  the indistinguishability of its constituents (stars or dark matter particles). This represents a
  refinement in the study of systems exhibiting incomplete violent relaxation. Within a
  combinatorial analysis similar to that of Lynden-Bell, we make use of this association to obtain
  a distribution function that deviates from the Maxwell-Boltzmann distribution, increasing its
  slope for high energies. Considering the smallness of the occupation numbers for large distances
  from the center of the system, we apply a correction to Stirling's approximation which increases
  the distribution slope also for low energies. The distribution function thus obtained presents
  some resemblance to the ``S'' shape of distributions associated with cuspy density profiles (as
  compared to the distribution function obtained from the Einasto profile), although it is not
  quite able to produce sharp cusps. We also argue how the association between mixing level and
  indistinguishability can provide a physical meaning to the assumption of particle-permutation
  symmetry in the N-particle distribution function, when it is used to derive the one-particle
  Vlasov equation, which raises doubts about the validity of this equation during violent
  relaxation.
\end{abstract}

\maketitle

\section{Introduction}
\label{sec:introduction}
Self-gravitating systems are known to present conceptual challenges for their description in terms
of thermodynamics and statistical mechanics, e.g. non-extensivity, negative heat capacity and the
inequivalence of (or even the impossibility of defining) canonical and microcanonical ensembles -
see \cite{Padmanabhan_1990,Binney_2008}. The main source of these difficulties lies in the
long-range nature of the gravitational interaction: differently from an ideal molecular gas in
which particles remain in uniform motion only modified by close-encounters, in self-gravitating
systems the particles (e.g. stars or dark matter constituents) are always interacting with the
gravitational field collectively produced. Also, differently from charged plasmas, in
self-gravitating systems the interaction is only attractive and there is no shortening of the
interaction range such as the Debye shielding. As a consequence of gravitational instability,
density contrasts tend to increase, leading to the appearance of non-linear phenomena that cannot
be treated perturbatively.

From the observational point of view, the common shape of many elliptical galaxies seems to
represent a final equilibrium configuration, despite the fact that the relaxation time for two-body
processes is larger than the age of the Universe \cite{Binney_2008}. The process that can explain
this relaxed state is violent relaxation: particles attain a quasi-stationary state by interacting
with the violently changing gravitational field during the first stages of structure collapse
\cite{Henon_1964, King_1966}. The time-scale for this process is the crossing time $\tau_{cr}$, the
time necessary for a particle to cross the galaxy, which is much lower than $\tau_{col}$, the
time-scale of relaxation by two-body, or collisional, processes \cite{LyndenBell_1967}. Thus, on
time-scales smaller than $\tau_{col}$, self-gravitating systems can be treated as collisionless,
i.e. without two-body interactions, in such a way that a test particle can be considered as only
interacting with the collectively generated mean gravitational field.

N-body simulations also provide important information about the stationary state achieved by these
systems after the collapse. For example the cuspy, ``universal'' density profiles of dark matter
halos \cite{NFW_1997} are well fit by simple functions, such as the NFW or Einasto profiles
\cite{Navarro_2004, Merritt_2005, Merritt_2006} - see Appendix
\ref{sec:f_eps_Einasto}. Interestingly, the observed projected density profiles of galaxy clusters
measured via gravitational lensing seem to be well fit by these same functions \cite{Umetsu_2011_b,
  Beraldo_Lima_Sodre_2013}. For galaxies, the situation is more complicated, as some of them (the
cored cases) are not well fit by these functions, presumably due to the influence of baryonic
components such as stars, gas, supernovae explosions etc \cite{2014Natur.506..171P}, or because of
a possible dark matter self-interaction \cite{Spergel_Steinhardt}. This is the so-called cusp-core
problem.

Despite the success of numerical simulations in reproducing some properties of the observed
objects, a clear explanation of the process driving a collisionless self-gravitating system to
equilibrium is lacking. Even globular clusters, classically viewed as being characterized by
collisional processes, seem to present evidences of collisionless dynamics \cite{Williams_2012},
which are yet to be clearly understood. See
\cite{LyndenBell_1967,Shu_1978,Madsen_1987,1987ApJ...316..502S} for some important papers on the
subject, \cite{Efthymiopoulos_2007,Bindoni_Secco_2008} for reviews and
\cite{Hjorth_Williams_2010_I,Pontzen_Governato_2013} for recent models.

For collisionless systems, it is usually assumed that the evolution of the one-particle
distribution function $f\PC{{\bf x,v},t}$ is governed by the Vlasov (or collisionless Boltzmann)
equation \citep[see][]{Binney_2008,Saslaw}
\begin{equation} \label{eq:Vlasov} \frac{df}{dt} = \frac{\partial f}{\partial t} + {\bf v} \cdot
  \frac{\partial f}{\partial {\bf r}} - \nabla \phi \cdot \frac{\partial f}{\partial {\bf v}} =
  0,
\end{equation}
where $\phi$ is the self-consistent, collectively generated gravitational potential.

As we discuss in \S\ref{sec_LB}, in a study of violent relaxation processes \cite{LyndenBell_1967},
Lynden-Bell translates the constraint provided by this equation into an exclusion principle when
maximizing the number of configurations (complexions) compatible with the conservation of energy
and total mass. In this procedure, it is assumed that the system is well mixed, i.e. that each
particle\footnote{There is an interesting discussion regarding the use of particles or phase
  elements (exploring the fluid analogy) - see \cite{Shu_1978}. Here we will just refer to
  particles.} has equal \emph{a priori} probability to be in any region of phase-space. This
hypothesis is known to be appropriate e.g. for ideal gases, for each molecule is able to assume any
position and velocity due to the highly random motions provided by collisions with other
molecules. For instance, in a gas in normal conditions of pressure and temperature, each molecule
suffers $\approx 10^5$ collisions per second. In some sense, we could say that each molecule
approximately occupies all available phase-space in a relatively small time-scale.  This is the
reason why one can assume that particles have equal \emph{a priori} probability to be in any region
of phase-space, thus allowing the equivalence between temporal and phase-space averages, the
so-called Ergodic Hypothesis \cite{Lichtenberg}. However, there are situations in which such mixing
is not complete (see \cite{Madsen_1987, Chavanis_2006} and references therein), in the sense that
the particles are not able to visit all regions of phase-space, particularly in self-gravitating
systems at the end of violent relaxation. This exposes the need for a model that deals with
intermediate mixing levels.

In this work, we propose a connection between the mixing level and the concept of
indistinguishability, and study the implications of this association for the quasi-stationary
states generated by the violent relaxation process. By ``mixing'' we do not mean ``phase mixing'',
which is a process associated to deterministic orbits in an integrable potential. Instead, we refer
to ``chaotic mixing'', i.e. that related to the exponential divergence of stochastic trajectories,
that allows each particle to explore a large region of phase-space and consequently different
particles to visit the same regions of phase-space. See \cite{2005NYASA1045....3M} for an overview
of this distinction and for references to important works on these lines. Although we did not make
this quantitative analysis, it would be possible to estimate this mixing level in N-body
simulations comparing temporal and phase-space averages in different regions of phase-space, for
example. As we will see in \S\ref{sec_part_mixed_dist_func}, in this work we used a very simplified
criterion to classify well mixed and poorly mixed regions.

We start in \S\ref{sec_LB} by describing how to obtain the Lynden-Bell distribution from
combinatorial arguments, making explicit the role of the distinguishability. In
\S\ref{sec_dist_indist} we discuss the concept of indistinguishability and present a criterion to
define it in terms of mixing.  In \S\ref{sec_part_mixed_dist_func} we determine a new distribution
function obtained according to this criterion and calculate the density profile $\rho(r)$ generated
by this distribution. Similar to the Isothermal Sphere, this density profile yields infinite mass
due to scaling in the external regions. As a solution to this problem we take into account the
smallness of occupation numbers in this region, as proposed by \cite{Hjorth_Williams_2010_I}. In
\S\ref{sec_small_numbers} we introduce this correction and present the resulting distribution
function and density profile.
In \S\ref{sec:Vlasov} we show how the criterion proposed gives a physical interpretation to the
hypothesis of permutation symmetry of the N-particle distribution function, which is assumed in
deducing the Vlasov equation by means of the BBGKY hierarchy. We argue that this symmetry
hypothesis, and consequently the Vlasov equation, may not be valid during violent
relaxation. Finally, in \S\ref{sec_conclusion} we summarize our results and discuss possible tests
of this model.

\section{Lynden-Bell distribution function}
\label{sec_LB}
The most important feature of Lynden-Bell's statistical analysis of violent relaxation is the
introduction of an exclusion principle due to the constraint imposed by the Vlasov equation,
Eq.~(\ref{eq:Vlasov}). Since in this case the phase-space density is constant, it is argued that
each particle occupies its own region in phase-space (its own micro-cell), without superposition
with other regions.

In order to obtain the distribution function from a combinatorial analysis, we divide the
phase-space into $J$ macro-cells \citep[see][]{LyndenBell_1967, Shu_1978}. Each macro-cell $i$ is
divided into $\nu_i$ micro-cells, of which $n_i$ are occupied by one particle and the other $\nu_i
- n_i$ micro-cells are empty. For simplicity, we consider that all the particles have the same mass
$m$. In the case of the simplest models of dark matter particles, this is exactly what is expected,
but in the case of stars in globular clusters or galaxies a mass distribution could bring some
differences - see \cite{LyndenBell_1967,Shu_1978}. In this way the total mass of the system is
\begin{equation}
  M = \displaystyle \sum_{i = 1}^J n_im = Nm,
\end{equation}
where $N$ is the total number of particles. The objective of the following calculation is to derive
the distribution function $F$, that represents the average number of particles per state ($F\propto
n_i/\nu_i$), maximizing the number of complexions, i.e. the number of micro-sates $W(\{n_i\})$
compatible with the macroscopic constraints of energy and mass conservation. The total energy is
given by
\begin{equation}
H = \displaystyle \sum_{i = 1}^J n_im\PC{\frac{1}{2}|{\bf v}_i|^2 + \frac{1}{2}\phi_i},
\end{equation}
where
\begin{equation}
\phi_i = \displaystyle - \sum_{j = 1, j\neq i}^J \frac{G m n_j}{|{\bf x}_i - {\bf x}_j|}
\end{equation}
is the potential in the $i$-th macro-cell, with position and velocity represented by ${\bf x}_i$
and ${\bf v}_i$ respectively. The calculation of $W$ involves two steps: determining the number of
possible configurations inside a macro-cell and the number of possibilities for exchanges between
different macro-cells.

Inside the $i$-th macro-cell, the number of ways to organize $n_i$ \emph{distinguishable} particles
in $\nu_i$ available micro-cells, but no more than one particle per micro-cell, is
\begin{equation}
\omega_i = \frac{\nu_i !}{\PC{\nu_i - n_i}!}.
\label{eq_configs_in_macrocell_dist}
\end{equation}
The same happens for all macro-cells $i = 1, 2, ...,J$, and so the total number of possibilities
for exchanges inside macro-cells is $\omega_1\cdot\omega_2\cdots \omega_J$.

For exchanges between different macro-cells, the number of ways to organize $N$
\emph{distinguishable} particles in the $J$ macro-cells, keeping fixed the number $n_i$ of
particles in each macro-cell is
\begin{equation}
\mathcal{N}_J = \frac{N!}{n_1 ! \cdot n_2 ! \cdots n_J !},
\label{eq_configs_between_macrocells}
\end{equation}
and the total number of complexions is
\begin{equation}
  W(\{n_i\}) = \PR{\frac{N!}{n_1 ! ... n_J !}}\PR{\frac{\nu_1 !}{\PC{\nu_1 - n_1}!}... \frac{\nu_J !}{\PC{\nu_J - n_J}!}}.
\label{eq_W_dist}
\end{equation}
To obtain the equilibrium configuration, we maximize the entropy $S=\ln W$ with respect to the
occupation numbers $n_i$, introducing the constraints of mass and energy conservation with Lagrange
multipliers $\lambda$ and $\eta$, which implies
\begin{equation}
  \delta \ln W - \lambda \delta M - \eta \delta H  = 0.
\end{equation}
Now we use Stirling's approximation
\begin{equation}
  \label{eq:stirling}
  \ln n! \approx n\PC{\ln n - 1},
\end{equation}
which is valid for $n\gg 1$. Note, however, that in the external regions of self-gravitating
systems, where the density goes to zero, this approximation is not expected to be valid, as noticed
by \cite{Hjorth_Williams_2010_I}. Neglecting momentarily this point and using
Eq.~(\ref{eq:stirling}), we obtain
\begin{equation}
\ln \PC{\frac{\nu_i - n_i}{n_i}} = \lambda m + \eta m E_i,
\label{eq_interm_dist}
\end{equation}
where
\begin{equation}
E_i = \frac{1}{2}|{\bf v}_i|^2 + \phi_i < 0
\label{eq_energy_macro_cell}
\end{equation}
is the energy per unit mass of the $i$-th macro-cell. Finally, we obtain the Lynden-Bell
distribution function
\begin{equation}
  \frac{n_i}{\nu_i} \propto \frac{f\PC{\varepsilon_i}}{f_0} = F\PC{\varepsilon_i} = \frac{1}{1 + e^{-\beta\PC{\varepsilon_i - \mu}}},
\label{eq_LB_dist_func}
\end{equation}
where $f_0$ is the fine-grained phase-space density, kept constant during all violent relaxation
process due to the constraint of Vlasov equation. This distribution is identical to the Fermi-Dirac
distribution, despite the use of distinguishable particles. In the above expression, we have
defined dimensionless energies as:
\begin{equation}
\varepsilon_i = -\frac{E_i}{|\phi_{i0}|} = \varphi_i - \frac{1}{2}{\bf u}_i^2,
\label{eq:energy_dimensionless}
\end{equation}
where $\varphi_i = -\phi_i/|\phi_{i0}|$ is the dimensionless gravitational potential, ${\bf u}_i =
{\bf v}_i/|\phi_{i0}|^{1/2}$ is the dimensionless velocity, $\phi_{i0} = \phi_i(0)$ is the central
potential and finally $\beta=\eta m |\phi_{i0}|$ and $\mu = \lambda/\eta|\phi_{i0}|$ are
dimensionless parameters. The parameter $\mu$, analogous to the chemical potential, determines the
position of the transition between two-regimes of small and high occupation numbers (degenerate
situation). The parameter $\beta$, analogous to the temperature, determines how abrupt this
transition is.

If, in order to guarantee the dynamical exclusion principle, we require that $n_i/\nu_i \ll 1$, or
$F\PC{\varepsilon_i} \ll 1$, we see that the distribution function (\ref{eq_LB_dist_func}) tends to
the Maxwell-Boltzmann case $F\PC{\varepsilon} = \exp\PR{\beta(\varepsilon - \mu)}$, that would be
obtained if we had not introduced the exclusion principle. Besides this conceptual problem, we know
that the Maxwell-Boltzmann distribution yields an infinite mass system, which contradicts the
assumption of finite mass \cite{Shu_1978}. Another criticism to Lynden-Bell's approach is that it
assumes equiprobability of all micro-states, but as violent relaxation occurs in such a short time
scale, possibly there is not enough time to complete the mixing process (see \cite{Madsen_1987,
  Chavanis_2006} and references therein).

Let us return to the calculation of $W$, but now treating particles as indistinguishable
\cite{Kull_Treuman_Bohringer_1996}. The number of ways to organize $n_i$ \emph{indistinguishable}
particles in $\nu_i$ micro-cells, instead of Eq.(\ref{eq_configs_in_macrocell_dist}), is
\begin{equation}
\omega_i = \frac{\nu_i !}{n_i !\PC{\nu_i - n_i}!}.
\label{eq_configs_in_macrocell_indist}
\end{equation}
Now with \emph{indistinguishable} particles, the exchange between different macro-cells, keeping
the number $n_i$ of particles per macro-cell fixed, does not produce different micro-states and now
we have $\mathcal{N}_J = 1 \Rightarrow$
\begin{equation}
  W(\{n_i\}) = \frac{\nu_1 !}{n_1 !\PC{\nu_1 - n_1}!}\cdots \frac{\nu_J !}{n_J !\PC{\nu_J - n_J}!}.
\label{eq_W_indist}
\end{equation}
But since the only difference from Eq.~(\ref{eq_W_dist}) obtained with distinguishable particles is
the factor $N!$, and for the maximization only the occupation numbers $n_i$ are relevant, the final
distribution function is exactly the same as Eq. (\ref{eq_LB_dist_func}).

At this point, one can argue that the above results indicate the unimportance of 
(in)distinguishability in the derivation of the distribution function. However, in
\S\ref{sec_part_mixed_dist_func} we propose that indistinguishability must be associated to the
mixing level of the system. According to this criterion, the scheme of
\cite{Kull_Treuman_Bohringer_1996} seems consistent because it assumes indistinguishable particles
and complete mixing (equiprobability of states). On the other hand, the scheme of
\cite{LyndenBell_1967} seems inconsistent, because it assumes 
equiprobability while taking distinguishable particles.

\section{Particle Indistinguishability}
\label{sec_dist_indist}
As discussed in the previous section, in the statistical interpretation of entropy formulated by
Boltzmann, the most probable thermodynamic states (macro-states) are those with the largest number
of micro-states compatible with the constraints of the problem, i.e. the largest number of
complexions $W$. In counting these states, the distinguishability is conceptually important because
one needs to know whether the permutation between two particles characterizes a new
micro-state. The particles are called distinguishable when this permutation creates a new
micro-state and indistinguishable when the permutation does not create a new micro-state.

According to the standard picture, as found in textbooks \cite{Tolman,1987stme.book.....H},
identical particles must be treated as indistinguishable in the context of quantum mechanics (due
to the superposition of wave functions) and as distinguishable in the context of classical
mechanics \citep[see][]{Tolman}. In this respect, the ideal gas was originally treated as being
constituted of distinguishable particles. Later it was realized that this assumption leads to
undesirable consequences such as the Gibbs paradox (see Appendix \ref{sec:Gibbs}) and required an
\textit{ad hoc} correction equivalent to treating the system as consisting of indistinguishable
particles. With the advent of quantum mechanics, this solution has been considered definitive,
because the gas particles should ultimately have quantum behavior and thus be indistinguishable
\cite{Tolman}. On the other hand, in the ideal crystal model, particles are treated as
distinguishable \cite{Tolman, Landau_stat_mec}. The common justification is that each particle is
confined to a well-defined region of space, oscillating around an equilibrium point without
superposing the wave functions of neighbor molecules.

Thus, it is commonly accepted that indistinguishability is only justified in the presence of
quantum effects and that in the absence of such effects, particles have to be treated as
distinguishable. However, it has been shown many years ago that it is perfectly possible to
formulate a statistical mechanics of indistinguishable particles in the context of classical
mechanics \cite{Schonberg_I, Schonberg_II}. Also, it is intriguing that when studying colloids
(systems composed of particles of intermediate size between large molecules and small grains in
suspension, i.e. macroscopic particles), the use of standard expressions for the entropy with the
assumption of distinguishable particles leads to the same conceptual contradictions of the ideal
gas of distinguishable particles \cite{Swendsen_2006}.

Therefore, a universal criterion for defining particle (in)distinguishability does not seem to be a
trivial issue \citep[see][]{Muynck_Liempd_1986}. Before presenting our proposed criterion, we note
that in the study of N-body dynamical systems it is common to observe the presence of separated
regions (``islands'') of phase-space inside which particles are mixed, i.e. continuously filling
the phase-space with stochastic trajectories, but not mixed to particles in other islands
\cite{Lichtenberg}.

With this picture in mind, we propose that particles in a mixed region of phase-space must be
treated as indistinguishable among themselves, but distinguishable from particles in a different
island. This criterion has some similarity with that discussed by \cite{Versteegh_2011}, according
to which the kind of permutations that are important to distinguishability are not mere changes of
index, but those that can really (physically) be performed. In this sense, the permutation of two
identical particles in a region of phase-space accessible to both does not create a new
micro-state, thus particles should be treated as indistinguishable. However, if these particles are
each one in a different region of phase-space, mutually inaccessible to each other, a permutation
represents a new micro-state and particles should be treated as distinguishable.

Contrary to the standard scenario, the criterion proposed here allows us to treat systems (under
certain circumstances) as composed by indistinguishable particles even if these components are
macroscopic objects like colloidal particles or stars\footnote{It is worth mentioning that Saslaw
  \cite{Saslaw_1969}, 45 years ago, had also discussed the possibility of a parametrization of
  levels of distinguishability in gravitational systems, but with an approach different from
  ours.}. The relation between this proposed criterion and the incompleteness of violent relaxation
will be discussed in the next section, where we determine the resulting distribution function. In
\S\ref{sec:Vlasov} we apply this reasoning to argue that the Vlasov equation may not be valid
during violent relaxation.

\section{Partially mixed distribution functions}
\label{sec_part_mixed_dist_func}
There are evidences that the violent relaxation process in self-gravitating systems is not able to
produce full mixing in phase-space \cite{Chavanis_2006, Bindoni_Secco_2008, Teles_2010,
  Teles_2011}, i.e. particles cannot access all possible micro-states before the achievement of a
stationary state. With this in mind and following the association discussed in
\S\ref{sec_dist_indist}, we make a combinatorial analysis similar to Lynden-Bell's scheme but
treating particles as indistinguishable for exchanges inside well mixed regions (in phase-space),
but distinguishable for exchanges between disconnected regions, i.e. not mixed together.

Using numerical simulations, \cite{Kandrup_1993} have concluded that during violent relaxation,
despite particles forgetting their initial positions and velocities, the ordering of the particles
energies is approximately conserved during the evolution of the system. In some sense, this is
equivalent to particles with similar energies being mixed among them but not with particles of
different energies. Since the energy is defined in a coarse-grained sense for each macro-cell,
Eq.~(\ref{eq_energy_macro_cell}), we use the criterion proposed here to treat particles as
indistinguishable for exchanges inside a macro-cell but distinguishable for exchanges between
macro-cells. A more precise classification could be done defining some index measuring how randomic
is the energy ranking in respect to the initial energies. This index could be monitored in N-body
simulations, but this is out of the scope of the present work.

In our analysis, we do not use the Vlasov equation as a constraint translated into an exclusion
principle as done by Lynden-Bell. The first reason for this is the theoretical problem already
discussed in \S\ref{sec_LB}: requiring that ${F\PC{\varepsilon_i}\ll1}$ -- in order to guarantee
the exclusion principle -- leads to a Maxwell-Boltzmann distribution, which is exactly what would
be obtained without the exclusion principle. The second reason is due to the possible non-validity
of the Vlasov equation during violent relaxation, as discussed in \S\ref{sec:Vlasov}.

The distribution function is calculated as follows: the number of ways to organize $n_i$
\emph{indistinguishable} particles inside a macro-cell allowing co-habitation in the $\nu_i$
micro-cells is given by
\begin{equation}
\omega_i = \frac{\PC{n_i + \nu_i - 1} !}{n_i !\PC{\nu_i - 1}!},
\label{eq_configs_in_macrocell_indist_cohab}
\end{equation}
which is the same factor as in the Bose-Einstein distribution. Together with expression
(\ref{eq_configs_between_macrocells}) for exchanges of \emph{distinguishable} particles between
macro-cells, and neglecting unity terms, the number of complexions results
\begin{equation}
  W(\{n_i\}) = \PR{\frac{N!}{\PC{n_1 !}^2 \cdots \PC{n_J !}^2}}\PR{\frac{\PC{n_1 + \nu_1} !}{\PC{\nu_1}!}\cdots \frac{\PC{n_J + \nu_J} !}{\PC{\nu_J}!}}.
\label{eq_W_dist_indist_cohab}
\end{equation}
Now following the same procedures as before and maximizing $\ln W$ subject to energy and mass
conservation, instead of Eq. (\ref{eq_interm_dist}), we obtain
\begin{equation}
\ln \PC{\frac{\nu_i + n_i}{n_i^2}} = \lambda m + \eta m E_i,
\label{eq_interm_dist_indist_cohab}
\end{equation}
from which we finally have:
\begin{equation}
  F\PC{\varepsilon} = \frac{1}{2} e^{\beta(\varepsilon - \mu) - k}\PC{\sqrt{1 +
      4e^{-\beta(\varepsilon - \mu) + k}} + 1},
\label{eq_dist_func_dist_indist_II}
\end{equation}
where $k = \ln \nu_i$. Note that, differently from the Lynden-Bell or Maxwell-Boltzmann, this
distribution function depends on the number $\nu_i$ of micro-cells accessible inside each
macro-cell. In principle, we could suppose that this number has some dependence on energy, but here
we treat it as a constant, being a parameter degenerated with $\beta$ and $\mu$~\footnote{From now
  on, for simplicity we omit the indices in the variables and parameters of the distribution
  function.}.

\begin{figure}[h]
\includegraphics[scale=0.5]{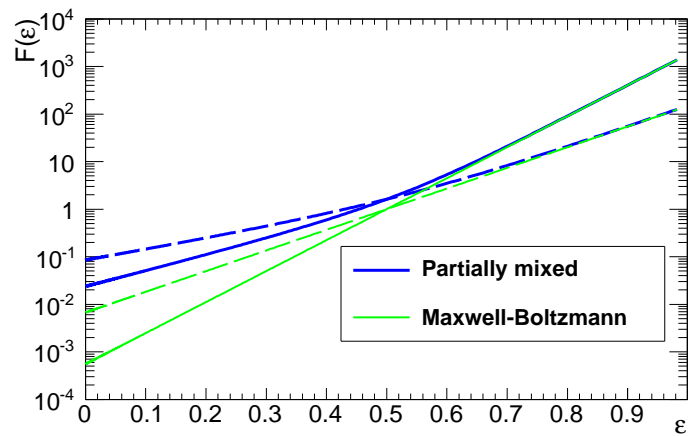}
\caption{Distribution function proposed in this work, Eq.~(\ref{eq_dist_func_dist_indist_II})
  (thick blue), in comparison with the Maxwell-Boltzmann distribution (thin green). The proposed
  distribution matches the Maxwell-Boltzmann for $\varepsilon > \mu$, but deviates to another
  Maxwell-Boltzmann with twice the original ``temperature'' for $\varepsilon \lesssim \mu$. All
  curves are for $\mu=0.5$. Continuous curves are for $\beta=15$ and dashed curves for $\beta=10$.}
\label{img_f_eps}
\end{figure}

This distribution function is shown as the thick blue lines in Fig.~(\ref{img_f_eps}) for
$\mu=0.5$, $\beta = (10, 15)$ and $k=0$. We see that it approaches the Maxwell-Boltzmann
distribution in the region $\varepsilon > \mu$, which represents the low velocity regime [see
Eq.~\eqref{eq:energy_dimensionless}]. On the other hand, for $\varepsilon \lesssim \mu$ (high
velocities), we have $F\PC{\varepsilon} \propto \exp\PR{\frac{1}{2}\beta(\varepsilon - \mu)}$,
which represents another Maxwell-Boltzmann distribution with twice the original ``temperature''.

\begin{figure}[h]
\includegraphics[scale=0.5]{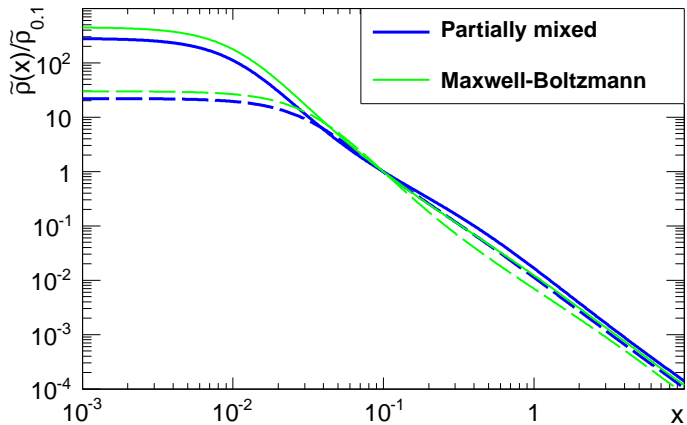}
\caption{Density profiles generated by our model (thick blue), in comparison with the Isothermal
  Sphere generated by the Maxwell-Boltzmann distribution (thin green). Values of parameters are the
  same as in Fig.~(\ref{img_f_eps}).}
\label{img_rho}
\end{figure}

Having determined the distribution function $F\PC{\varepsilon}$,
Eq.~\eqref{eq_dist_func_dist_indist_II}, we can now calculate the density profiles $\rho\PC{r}$ of
spherically symmetric and isotropic structures generated by the model. In order to do that, we
define a dimensionless distance from the center $x = r/a$, a density profile
$\tilde{\rho}=\rho/(f_0|\phi_0|^{3/2})$ and the constant $A= 4\pi G \sqrt{|\phi_0|}a^2 f_0$, where
$a$ is a scale parameter and $G$ is the gravitational constant. In these units, we have
\begin{equation}\label{eq:rho_from_f}
  \tilde{\rho}\PC{\varphi} = 4\pi\int_{-\infty}^\varphi F(\varepsilon) \sqrt{2(\varphi - \varepsilon)} \, \mathrm{d}\varepsilon.
\end{equation}
With this relation, we solve Poisson equation $({\nabla^2\varphi=-A\tilde{\rho}})$ to determine
$\varphi(x)$ and consequently $\tilde{\rho}(x)$. Supposing spherical symmetry, this equation reads
\begin{equation}
\frac{d^2\varphi}{dx^2} + \frac{2}{x}\frac{d\varphi}{dx} = - A\tilde{\rho}\PC{\varphi}.
\end{equation}
We numerically solve this equation with a 4-th order Runge-Kutta algorithm imposing the conditions
$\varphi(0) = 1$ and ${d\varphi(0)}/{dx} = 0$ and fixing $A=10$. The results are shown in
Fig. (\ref{img_rho}), with the density profiles normalized by their values at $x=0.1$.

\begin{figure}
  \centering
  \includegraphics[scale=0.5]{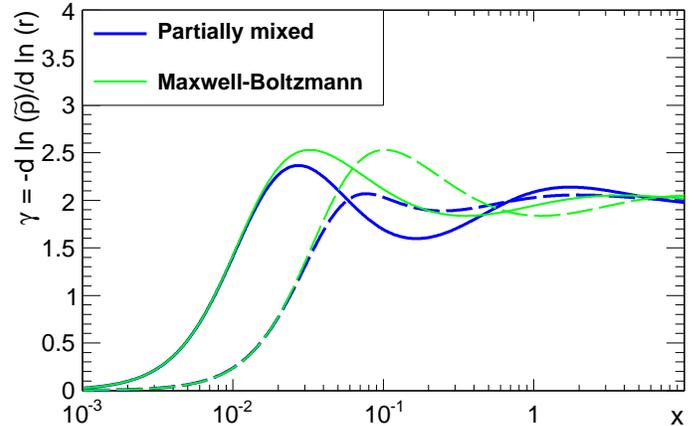}
  \caption{Density slope of the model proposed here (thick blue lines) in comparison with that
    generated by the Maxwell-Boltzmann distribution (thin green lines). The slope goes to 2 in the
    external regions, what produces an infinite mass. Parameters as in Fig.~(\ref{img_f_eps}).}
  \label{img_gamma}
\end{figure}

We see that our model generates a density profile similar to that of the Isothermal Sphere
generated by the Maxwell-Boltzmann distribution: it has a core and it oscillates around $\propto
x^{-2}$ in the external regions. This can be more clearly seen in Fig.~(\ref{img_gamma}), that
shows the density slope $\gamma(x) = -d\ln (\rho)/d\ln(x)$. As is well known, a density profile
varying as $\propto x^{-2}$ in the external region generates an infinite mass distribution, in
contradiction with the initial constraint of finite mass, and our model is not able, \emph{per se},
to solve this problem. Instead, we need an extra ingredient related to a correction for the
smallness of occupation numbers in the external region, which is discussed in the next section.

Although not used in the rest of the paper, we also consider the case of introducing an exclusion
principle, since it can be tested in other applications. The number of ways to organize
\emph{indistinguishable} particles inside a macro-cell, preventing co-habitation of micro-cells, is
given by expression (\ref{eq_configs_in_macrocell_indist}). Together with expression
(\ref{eq_configs_between_macrocells}) for the number of ways to exchange \emph{distinguishable}
particles between macro-cells, it results in
\begin{equation}
W(\{n_i\}) = \PR{\frac{N!}{\PC{n_1 !}^2 \cdots \PC{n_J !}^2}}\PR{\frac{\nu_1 !}{\PC{\nu_1 - n_1}!}\cdots \frac{\nu_J !}{\PC{\nu_J - n_J}!}}.
\label{eq_W_dist_indist}
\end{equation}
Following the same procedures as before, we obtain
\begin{equation}
\ln \PC{\frac{\nu_i - n_i}{n_i^2}} = \lambda m + \eta m E_i,
\end{equation}
from which results
\begin{equation}
  F\PC{\varepsilon} = \frac{1}{2} e^{\beta(\varepsilon - \mu) - k}\PC{\sqrt{1 +
      4e^{-\beta(\varepsilon - \mu) + k}} - 1}.
\label{eq_dist_func_dist_indist_I}
\end{equation}
It is interesting to note that, as in Fermi-Dirac versus Bose-Einstein distributions, the only
difference between Eqs. (\ref{eq_dist_func_dist_indist_II}) and (\ref{eq_dist_func_dist_indist_I})
is a $\pm$ sign.

\section{Correction for small occupation numbers}
\label{sec_small_numbers}
In the external regions, as the density profile goes to zero, the occupation numbers assume small
values, invalidating the use of the Stirling's approximation, Eq.~(\ref{eq:stirling}), in deriving
the distribution function. With this in mind, \cite{Hjorth_Williams_2010_I} proposed a correction
that when applied to the Maxwell-Boltzmann case, gives rise to a distribution function identical
to that of King models \cite{King_1966}, which goes smoothly to zero as $\varepsilon$ approaches a
free parameter $\varepsilon_0$.

The maximization procedure done in \S\ref{sec_LB} can be represented identifying $n! =
\Gamma(n+1)$ and remembering the definition of the digamma function $\psi(n) = d\ln \Gamma/dn$. The
Stirling's approximation, Eq.~(\ref{eq:stirling}), corresponds to ${\psi(n+1) \approx \ln n}$ and
the correction proposed by \cite{Hjorth_Williams_2010_I} is given by 
\begin{equation}
 \psi(n+1)\approx\ln (n + e^{-\gamma}),
\end{equation}
where ${\gamma = -\psi(1) \approx 0.57721566}$ is Euler's constant. This approximation turns out to
be excellent, even for very small numbers - see Fig.~(1) of \cite{Hjorth_Williams_2010_I}.

In fact, if we take this correction for the Maxwell-Boltzmann distribution, as done by
\cite{Hjorth_Williams_2010_I}, we obtain
\begin{equation}
\ln \PC{\frac{\nu_i}{n_i+e^{-\gamma}}} = \lambda m + \eta m E_i,  
\end{equation}
which implies that
\begin{equation}
F(\varepsilon) = \PR{e^{\beta(\varepsilon - \mu)} - e^{-k-\gamma}}\,.
\end{equation}
As $F(\varepsilon)$ necessarily goes to zero for some energy, we see that the correction for small numbers
already introduces a dependence on $\nu_i = e^k$, even in the Maxwell-Boltzmann case. If we now
impose that the distribution function is zero for $\varepsilon = \varepsilon_0$, we have
\begin{equation}
  F(\varepsilon) = e^{\beta(\varepsilon_0 - \mu)}\PR{e^{\beta(\varepsilon - \varepsilon_0)} - 1},
\end{equation}
which corresponds to the King model \cite{King_1966}.

\begin{figure}[h]
\includegraphics[scale=0.5]{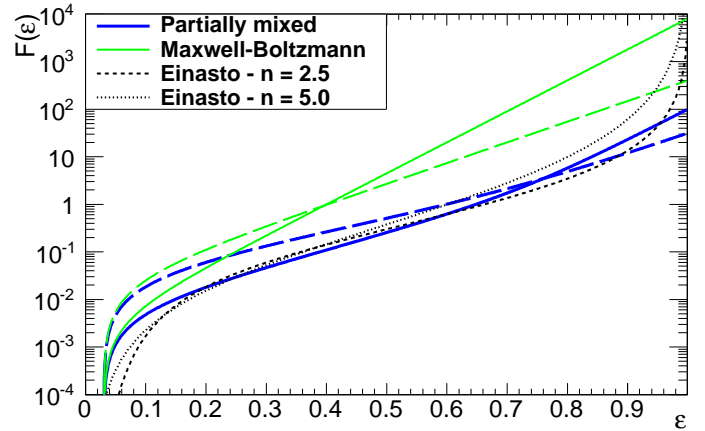}
\caption{Distribution function corrected for the smallness of occupation numbers,
  Eq.~(\ref{eq_f_eps_trunc}) (thick blue), in comparison with the Maxwell-Boltzmann (also
  corrected, equivalent to King's) distribution (thin green). Distribution functions
  associated with cuspy density profiles (like the Einasto, shown for two values of parameter $n$)
  have a ``S'' shape that our model is not quite able to reproduce, although the correction being
  in the right direction, i.e. increasing the slope of $F(\varepsilon)$ for high
  $\varepsilon$. All curves are for $\mu=0.4$ and $\varepsilon_0 = 0.03$. Continuous curves are for
  $\beta=15$ and big dashed curves for $\beta=10$.}
\label{img_f_eps_trunc}
\end{figure}

If we now apply this correction to our model, instead of Eq.~(\ref{eq_interm_dist_indist_cohab}),
we obtain
\begin{equation}
\ln \PC{\frac{\nu_i + n_i - 1 + e^{-\gamma}}{(n_i+e^{-\gamma})^2}} = \lambda m + \eta m E_i,  
\end{equation}
implying that
\begin{eqnarray*}
  F\PC{\varepsilon} = \frac{1}{e^{k+\gamma}}\times
\end{eqnarray*}
\begin{equation}
\chaves{\frac{1}{2} e^{\beta(\varepsilon - \mu) +\gamma}\PC{\sqrt{1 +
      4(e^k - 1)e^{-\beta(\varepsilon - \mu)}} + 1} - 1},
\label{eq_f_eps_trunc}
\end{equation}
and again doing $f(\varepsilon_0) = 0$ we have
\begin{equation}
  e^k = e^{-\gamma}\PC{e^{-\beta(\varepsilon_0 - \mu) - \gamma} - 1} + 1.
\end{equation}
In Fig.~(\ref{img_f_eps_trunc}), the thick blue lines show this distribution function for $\beta =
(10, 15)$, $\mu = 0.4$ and with $\varepsilon_0 = 0.03$. Again we see that for decreasing energies,
the distribution changes from a Maxwell-Boltzmann to another Maxwell-Boltzmann, but for
$\varepsilon \approx \varepsilon_0$ it goes to zero. The thin green lines are the King's models,
obtained as a Maxwell-Boltzmann corrected for small occupation numbers.

For a qualitative comparison, we also show the distribution function associated with the Einasto
density profile \cite{Einasto_1965} - see Appendix~\ref{sec:f_eps_Einasto}, which describes the
details of this calculation, made here for the first time, as far as we know. The two curves shown
are for $n=2.5$ and $n=5.0$, 
representing typical values for galactic and galaxy cluster scale respectively.

Note that the distribution function associated with the Einasto profile, as well as cuspy density
profiles in general \citep[see][]{Widrow_2000} has a ``S'' shape, with $F(\varepsilon)$ going to
zero for small $\varepsilon$ and going to increasing slopes for large $\varepsilon$. It is
interesting that the model proposed here, although not presenting exactly the same shape, gives a
correction in the same direction, increasing the slope for large values of $\varepsilon$.

\begin{figure}[h]
\includegraphics[scale=0.5]{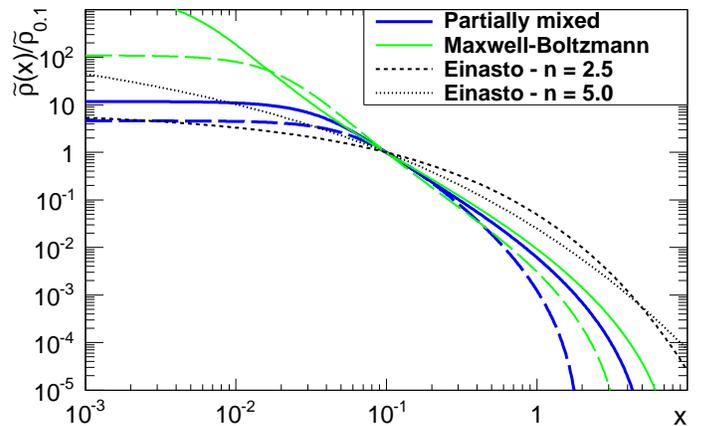}
\caption{Density profiles generated by the model corrected for small occupation numbers (thick
  blue), in comparison with that generated by the corrected Maxwell-Boltzmann distribution or
  King's model (thin green) and with the Einasto profile. Values or parameters are the same as in
  Fig.~(\ref{img_f_eps_trunc}).}
\label{img_rho_trunc}
\end{figure}

\begin{figure}
  \centering
  \includegraphics[scale=0.5]{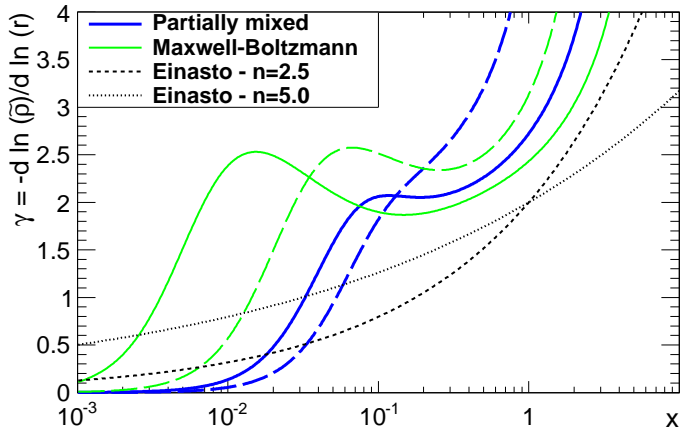}
  \caption{Density slope generated by the model corrected for small occupation numbers (thick blue)
    in comparison with that generated by the corrected Maxwell-Boltzmann distribution or King's
    model (thin green) and with the Einasto profile. Parameters as in
    Fig.~(\ref{img_f_eps_trunc}).}
  \label{img_gamma_trunc}
\end{figure}

As done previously, we calculate the density profiles generated by this function, which are shown
in Fig.~(\ref{img_rho_trunc}), again normalized by the density at $x=0.1$. We see that now the
density profile is steeper than $\propto x^{-2}$ in the external region, and in fact the problem of
infinite mass is solved. Also shown are the corrected Isothermal Sphere (King's model) and the Einasto
density profile.
For completeness, we also show in Fig.~(\ref{img_gamma_trunc}) the density slope obtained after the
correction for small occupation numbers.

\section{Validity of the Vlasov equation} \label{sec:Vlasov} Let us return to the discussion of the
violent relaxation process to see how the association proposed here (between the mixing level and
the indistinguishability) can give a clearer meaning to an important hypothesis assumed in
the deduction of the Vlasov equation.

Intuitively, one expects that during violent relaxation, a process that starts far from
equilibrium, the total field varies in a very complex way, producing chaotic motions and driving
the system to an equilibrium (or stationary) state. In fact, as pointed out in
\cite{Merritt_1996_2}, the presence of chaotic motions combined with the rapid approach to a
stationary state observed in numerical simulations seems to indicate this effect. However the
Vlasov equation is reversible in time, which is incompatible with a process driving the system to
an equilibrium (or relaxed) state characterized by a maximum entropy. In fact,
\cite{Tremaine_Henon_Lynden_Bell_1986} have shown that if the system is described by
Eq.~\eqref{eq:Vlasov}, there is no upper limit for the entropy associated with any convex function
$C(f)$, in particular for the Boltzmann entropy, represented by $C(f)=f\ln f$. The standard
argument to solve this problem is that the evolution to an equilibrium state is given in a
coarse-grained sense, while the Vlasov equation concerns the fine-grained distribution function.

The deduction of the equation governing the evolution of the one-particle distribution function $f$
is usually done starting from the Liouville equation. It states that an \emph{isolated} system
composed of $N$ particles collectively represented by the $N$-particle joint distribution function
$f^{(N)}\PC{\bf{x_1, p_1,...,x_N,p_N},t}$ necessarily respects \citep[see][]{Liboff}
\begin{equation}
  \label{eq:Liouville}
  \frac{df^{(N)}}{dt} = 0.
\end{equation}
This equation can be statistically interpreted as the evolution of the system as a whole being
smooth, free of sudden changes, which is adequate since it describes an isolated system (by
definition free of external influences), whose particles move according to Hamilton equations.

The next step to obtain the equation for the one-particle distribution $f$ is the construction of
the so-called BBGKY hierarchy \citep[see][]{Landau_PK,Binney_2008, Saslaw}, and it involves some
extra assumptions. The first one is the symmetry of $f^{(N)}\PC{\bf{x_1, p_1,...,x_N,p_N},t}$
relative to changes of coordinates and momenta of the particles. This makes the phase-space
averaged contribution of each particle to the total force exerted on the test particle to be the
same, implying Eq.~\eqref{eq:hyp_indist} - see Appendix~\ref{sec_BBGKY}. The second hypothesis is
that of molecular chaos, i.e., that the two-particle distribution function is just the product of
two one-particle distribution functions, the correlations being negligible, as is expressed by
Eq.~\eqref{eq:hyp_mol_chaos}. With these two hypothesis, one obtains Vlasov equation,
Eq.~\eqref{eq:Vlasov}.

Far from being just a calculation strategy, these assumptions have a deep statistical meaning, and
without them it is not possible (to the best of our knowledge) to obtain the Vlasov equation. The
symmetry of $f^{(N)}$ is commonly treated as a direct consequence of the assumption of
\emph{identical} particles \citep[see][]{Binney_2008, Saslaw}. However, there is no mechanical
principle that guarantees this symmetry. Instead, it is an extra hypothesis, with important
statistical content. In our context it is equivalent to treating particles not only as
\emph{identical} but as \emph{indistinguishable} and, according to the criterion discussed in
\S\ref{sec_dist_indist}, it refers to the possibility of all particles to visit the same regions
of phase-space. As already pointed out by \cite{Saslaw}, the BBGKY hierarchy was developed to
describe molecules in fluids and ions in plasmas close to equilibrium. These systems are very
different from a self-gravitating system during violent relaxation, which is a phenomenon that
starts far from equilibrium. Besides, as discussed in \S\ref{sec:introduction}, self-gravitating
systems are not able to fully mix the phase-space, or at least are expected to be much less
effective in doing so than plasmas or fluids in laboratory.

From this discussion, it seems that there is no reason to suppose that the symmetry of $f^{(N)}$ is
a valid hypothesis during violent relaxation. Besides that, the correlations in situations far from
equilibrium may be large \citep[see][]{Saslaw_2008}, which also may contradict the assumption of
molecular chaos, Eq.~(\ref{eq:hyp_mol_chaos}). Thus the Vlasov equation does not seem to be valid
during this process. In fact, as pointed out by \cite{NYAS:NYAS28}, the relation between the full
N-body problem and the associated transport equation (presumably the Vlasov equation) is far more
complicated than usually assumed. Although we do not propose any concrete alternative to Vlasov
equation, we expect that during violent relaxation the one-particle distribution $f$ must be
described by a full transport equation,
\begin{equation} \label{eq:Boltzmann} \frac{\partial f}{\partial t} + {\bf v} \cdot \frac{\partial
    f}{\partial {\bf r}} - \nabla \phi \cdot \frac{\partial f}{\partial {\bf v}} = L[f],
\end{equation}
where $L[f]$ is a stochastic term (see \cite{Kandrup_1980}) related to chaotic changes of the
potential.

This can be so even if we can neglect 2-body interactions. The idea of the ``collisional'' term in
the Boltzmann equation is associated, not exactly with collisions, but to any stochastic process that
suddenly changes the probability flux of the test particle \citep[see][]{Landau_PK}. In ideal
gases, these processes are realized by the collisions, but for systems with long-range
interactions, it can be a chaotically time changing mean field.  As the distribution function $f$
carries information about the system as a whole, the long-range interactions transmit the small
perturbations of all particles to the test particle.

The above discussion is relevant for what we have proposed in this paper for two reasons: first, it
reinforces our expectation that self-gravitating systems can achieve a stationary state as a
consequence of going through irreversible processes, without the need to advocate the
coarse-grained sense usually attributed to this evolution
\cite{Tremaine_Henon_Lynden_Bell_1986}. In other words, a real ``arrow of time'' is introduced if
the Vlasov equation is not valid during the violent relaxation process and the stationary state can
be predicted by the old strategy of maximizing the complexions, i.e. the number of possible
micro-configurations given the constraints of the problem. Secondly, it is questionable whether or
not the Vlasov equation should imply an exclusion principle constraint (as discussed in
\S\ref{sec_LB}).  In case this equation is not valid during violent relaxation, there would be even
less reason to impose such constraint. That is why we did not impose in our analysis the exclusion
principle proposed by Lynden-Bell \cite[][]{LyndenBell_1967}.

\section{Conclusion}
\label{sec_conclusion}

We propose a new criterion to choose between distinguishability and indistinguishability, namely
the level of mixing in phase-space. According to this, in systems that do not have enough time to
completely mix phase elements, particles in well mixed regions must be treated as indistinguishable
and particles in poorly mixed regions must be treated as distinguishable. This criterion is
consistent with the standard classification of ideal gases as being made of indistinguishable
particles and of crystals as being made of distinguishable particles. However it opens new
perspectives on the classification of systems of macroscopic constituents like colloids and
self-gravitating systems. It also provides a solution to the Gibbs paradox without the need of
arguments related to the quantum nature of particles (see Appendix \ref{sec:Gibbs}).

Violent relaxation is known to be incomplete, in the sense that the particles (or stars) cannot
explore all regions of phase-space before the achievement of a stationary state. Thus the model
proposed here can represent a solution to this problem, explicitly translating this incompleteness
in the combinatorial analysis. According to \cite{Kandrup_1993}, during violent relaxation the
particles ``forget'' the positions and velocities but keep the initial order in their energies. We
express this fact treating particles as indistinguishable for exchanges inside a macro-cell (that
defines an energy value) but as distinguishable for exchanges between macro-cells.

The result is a new distribution function that tends to the Maxwell-Boltzmann distribution for high
energies, but deviates to another Maxwell-Boltzmann distribution with twice the
original ``temperature'' for low energies.

The density profiles generated by this distribution function resemble those predicted by the
Maxwell-Boltzmann distribution, the Isothermal Sphere. As so, they vary as $\propto x^{-2}$ in the
external regions, yielding an infinite mass system. However, in the external regions the occupation
numbers are small, invalidating the use of Stirling's approximation. Using the correction proposed
by \cite{Hjorth_Williams_2010_I}, we obtain a distribution function that goes to zero for an energy
fixed by a parameter $\varepsilon_0$. The density profile generated by this corrected distribution
function is steeper in the external region, effectively solving the infinite mass problem. 

It is interesting to note that the corrected distribution function resembles the ``S'' shape of the
distribution associated with the Einasto profile, or at least provides a correction in the right
direction. The high energy part of this function determines the behavior of the density profile in
the inner region \cite{Widrow_2000} and our model seems to go in the right direction to generate
high densities that could mimic a cuspy profile.

The new distribution function obtained in this work can be tested and used in several astrophysical
applications. For example, the resulting density profile can be fit to gravitational lensing data
from galaxy clusters, as done for other models in \cite{Beraldo_Lima_Sodre_2013}. Another direct
test can be made with data from rotation curves of spiral galaxies. On smaller scales, analyses of
the density profile can also be made from optical data of globular clusters. A caveat that we
should point out is that the dissipative nature of the baryon collapse also affects the density
profile of astrophysical objects \cite{2012MNRAS.422.3081M}, which complicates the comparison with
observational data.

The velocity distribution associated to the model can be tested by fitting to the simulated data of
structure formation, as done by \cite{Vogelsberger_2009,Lisanti_2011} to test different models. In
the analysis of direct detection experiments such as the XENON \cite{Xenon_2012} and CDMS
\cite{CDMS_2014} projects, the velocity distribution is an important ingredient for the predicted
event rate associated to different dark matter particles and the model proposed here can be of some
utility in this context. Another possible application is in mass modeling methods such as MAMPOSSt
\cite{Mamon_2013}, in which the distribution of tracers in projected phase space is used to
determine the total density and anisotropy profiles, starting from the assumption of some 3D
velocity distribution.

We also considered the possibility of preventing co-habitation in micro-cells keeping the same
criterion for distinguishability as before, obtaining another distribution function, differing from
the first by a $\pm$ sign. Both distribution functions obtained here can be of some
importance for other phenomena far from equilibrium.

Finally, we showed how the association between the mixing level and indistinguishability gives a
physical meaning for the assumption of symmetry of the N-particle distribution function
$f^{(N)}$. This assumption is equivalent to treat particles as \emph{indistinguishable}, which is
a stronger assumption than treating them as \emph{identical} and, according to the criterion proposed here, it is
equivalent to the assumption that all particles have access to the same regions of phase-space,
i.e. that the system is completely mixed. Since it is well known that such mixing is not complete
in self-gravitating systems during violent relaxation, this suggests that the symmetry hypothesis is not
adequate and thus the Vlasov equation is not necessarily valid during this process.

\appendix

\section{Einasto distribution function}
\label{sec:f_eps_Einasto}
The Einasto density profile was proposed in the 60's to describe the surface brightness of
elliptical galaxies and recently has been used to fit data of simulated \cite{Navarro_2004,
  Merritt_2005} and observed structures in the galactic and galaxy cluster scales
\cite{Beraldo_Lima_Sodre_2013}, giving results better than NFW in many cases. In fact, it can mimic
the cuspy density profile described by NFW, despite being finite at the origin. The profile is
given by
\begin{equation}
  \label{eq:rho_Einasto}
  \tilde{\rho}(x)=\exp\chaves{-2n(x^{1/n} - 1)}.
\end{equation}
The associated potential is given by \cite{Montenegro_2012}:
\begin{eqnarray*}
  \varphi(x) = -\frac{\phi(x)}{|\phi(0)|} = \frac{\Gamma(3n)}{\Gamma(2n)}\frac{1}{(2n)^n x}\times
\end{eqnarray*}
\begin{equation}\label{eq:psi_Einasto}
  \left[1 - \frac{\Gamma(3n, 2nx^{1/n})}{\Gamma(3n)} + (2n)^nx\frac{\Gamma(2n, 2nx^{1/n})}{\Gamma(3n)}\right],
\end{equation}
where $\Gamma(a,x)$ is the (upper) incomplete gamma function. The distribution function
$F(\varepsilon)$ can be obtained from Eq.~(\ref{eq:rho_from_f}) through an Abel transform
\citep[see][]{Binney_2008}:
\begin{equation}
  F(\varepsilon) = \frac{1}{\sqrt{8}\pi^2}\left[ \int_0^\varepsilon \,
    \frac{d^2\tilde{\rho}}{d\varphi^2}\frac{d\varphi}{\sqrt{\varepsilon - \varphi}} +
    \frac{1}{\sqrt{\varepsilon}}\left( \frac{d\tilde{\rho}}{d\varphi}\right)_{\varphi=0}\right],
\end{equation}
where the second term in the square bracket is zero for the Einasto profile. Following
\cite{Widrow_2000}, we use Eqs.~(\ref{eq:rho_Einasto}) and (\ref{eq:psi_Einasto}) and solve the
integral above numerically. The results are shown in Fig.~(\ref{img_f_eps_trunc}) for two values of
$n$.

\section{The Gibbs paradox}
\label{sec:Gibbs}

Since the beginning of the development of statistical mechanics up to the present days, the Gibbs
paradox has brought many discussions and attempts of solution, definite for ones and unsatisfactory
for others. It can be described in various forms, in particular as follows \cite{Jaynes_1992}: an
enclosure volume $V$ is divided by a wall into two volumes $V_1$ and $V_2$ filled respectively with
$N_1$ and $N_2$ molecules with mass $m$ of the same gas subject to the same conditions of pressure
and temperature. Assuming that the particles are distinguishable, the initial number of
micro-states compatible with the macro-state 1, the same being for 2, is:
\begin{equation}
W_I^1 = \int d^3\vec{r_1}...\int d^3\vec{r_{N_1}}\int d^3\vec{p_1}...\int d^3\vec{p_{N_1}}\Rightarrow
\end{equation}
\begin{equation}
W_I^1 = V_1^{N_1}\cdot \frac{\PC{2\pi m}^\frac{3N_1}{2}}{\Gamma\PC{\frac{3N_1}{2}}}E_1^{\frac{3N_1}{2}}
\end{equation}

We calculate the entropy as $S_I^1 = \ln W_I^1$ (the same for 2) and the total entropy is $S_I =
S_I^1 + S_I^2$. Setting $f=N_1/N = V_1/V = E_1/E$, and applying the thermodynamic limit
$\PC{N\rightarrow\infty}$, the entropy per particle is
\begin{eqnarray*}
\frac{S_I}{N} = s_I = f\ln f + \PC{1 - f}\ln\PC{1 - f} +
\end{eqnarray*}
\begin{equation}
+\ln V + \frac{3}{2}\ln u + \frac{3}{2}\ln\PC{\frac{4\pi}{3}m} + \frac{3}{2},
\end{equation}
where $u = E/N$ is the energy per particle and it was used the Stirling's approximation, $\ln N!
\approx N\ln N - N$.

After removing the wall and considering the system as whole, recalculation of the total number of
micro-states gives
\begin{equation}
W_F = V^N\cdot \frac{\PC{2\pi m}^\frac{3N}{2}}{\Gamma\PC{\frac{3N}{2}}}E^{\frac{3N}{2}},
\end{equation}
and it follows that
\begin{equation}
s_F = \ln V + \frac{3}{2}\ln u + \frac{3}{2}\ln\PC{\frac{4\pi}{3}m} + \frac{3}{2}.
\label{eq_entropia_dist}
\end{equation}
Thus, the entropy before and after removing the division differ by
\begin{equation}
\Delta s = -\PR{f\ln f + \PC{1 - f}\ln\PC{1 - f}}.
\end{equation}
In the particular case of equal volumes we have
\begin{equation}
\Delta s = \ln 2.
\end{equation}
Therefore, if we consider that the gas particles are distinguishable, as classically assumed, we
conclude that there was an increase in entropy, which does not agree with the fact that there is no
macroscopic change in the system. Hence the paradox.

If on the other hand we assume that the particles are indistinguishable, we have to divide the
number of complexions obtained previously by the number of permutations of the $N_1$, $N_2$ or $N$
particles. Thus,
\begin{equation}
W_I^1 = \frac{1}{N_1!}\int d^3\vec{r_1}...\int d^3\vec{r_{N_1}}\int d^3\vec{p_1}...\int d^3\vec{p_{N_1}},
\end{equation}
the same for 2. Following the same steps as in the previous case, we arrive at
\begin{equation}
s_I = \ln v + \frac{3}{2}\ln u + \frac{3}{2}\ln\PC{\frac{4\pi}{3}m} + \frac{5}{2},
\end{equation}
where $v = V/N$. Again, removing the wall, the phase space and the entropy become
\begin{equation}
W_F = \frac{1}{N!}\int d^3\vec{r_1}...\int d^3\vec{r_{N}}\int d^3\vec{p_1}...\int d^3\vec{p_{N}}
\end{equation}
and
\begin{equation}
s_F = \ln v + \frac{3}{2}\ln u + \frac{3}{2}\ln\PC{\frac{4\pi}{3}m} + \frac{5}{2}\Rightarrow
\label{eq_entropia_indist}
\end{equation}
\begin{equation}
\Delta s = 0.
\end{equation}

Thus, removal of the wall does not produce increase in entropy (which is within our expectations),
once we consider that the particles are indistinguishable, the justification being traditionally
associated with the quantum behavior of the gas molecules.

On the other hand, according to this standard interpretation, a system with particles almost
identical (with arbitrarily similar mass, for example) would still present the same increase in
entropy. This discontinuity, passing from identical to arbitrarily similar particles, still cannot
be explained if the indistinguishability depends on the particles being strictly identical.

Furthermore, appealing to quantum mechanics to solve Gibbs paradox is deemed unsatisfactory to some
authors \cite{Versteegh_2011}, because it should not be necessary to use quantum arguments in a
strictly classical conceptual problem. In other terms, it is not an experimental evidence of need
for quantum physics, but a conceptual inconsistency internal to classical physics, which should be
solved in classical terms.

\section{BBGKY Hierarchy}
\label{sec_BBGKY}
Let a system with $N$-particles be described by the Hamiltonian
\begin{equation}
  H = \displaystyle \sum_{a = 1}^N \PR{\frac{{\bf p}_a^2}{2m} + \displaystyle \sum_{b=a+1}^N U\PC{|{\bf
        r}_a - {\bf r}_b|}},
\end{equation}
where $U\PC{|{\bf r}_a - {\bf r}_b|}$ is the potential energy. The evolution of the $N-$particle
distribution function $f^{(N)}$ is governed by Liouville equation [Eq.~(\ref{eq:Liouville})]. With
the help of Hamilton equations, it reads
\begin{equation}
  \frac{\partial f^{(N)}}{\partial t} + \displaystyle \sum_{a = 1}^N \PC{\frac{\partial
      f^{(N)}}{\partial{\bf r}_a} \frac{{\bf p}_a}{m} - \frac{\partial
      f^{(N)}}{\partial{\bf p}_a}\displaystyle \sum_{b=a+1}^N \frac{\partial U_{ab}}{\partial {\bf
        r}_a}} = 0,
\label{eq:Liouville_2}
\end{equation}
where $U_{ab} = U\PC{|{\bf r}_a - {\bf r}_b|}$.

We want to derive the equation for the one-particle distribution $f$. This function is not to be
interpreted as describing the evolution of some specific particle, but that of a test particle
randomly chosen. In this sense, it keeps statistical information about the system as a whole,
even-though referring to coordinates of one particle. Also we normalize the one-particle
distribution $f$ taking into account all the permutations between the $(N-1)$ remaining particles
\cite{Liboff}:
\begin{equation}
  f = \frac{N!}{(N-1)!}\int f^{(N)}(\Gamma_1,\dots,\Gamma_N)d\Gamma_2\dots d\Gamma_N,
\label{eq:def_f}
\end{equation}
where $d\Gamma_i = d{\bf r}_i d{\bf p}_i$. Following standard steps we integrate
Eq.~(\ref{eq:Liouville_2}) in $d\Gamma_2\dots d\Gamma_N$, obtaining
\begin{eqnarray*}
\frac{(N-1)!}{N!} \PC{\frac{\partial f}{\partial t} + \frac{\partial
      f}{\partial{\bf r}_1} \frac{{\bf p}_1}{m}} =
\end{eqnarray*}
\begin{equation}
=  \frac{\partial}{\partial {\bf p}_1}
  \displaystyle \sum_{b=2}^N \int f^{(N)} \frac{\partial U_{1b}}{\partial {\bf r}_1}d\Gamma_2\dots d\Gamma_N.
\end{equation}
The integral in the right hand side represents the force exerted on the test particle by each one
of the other particles, averaged over the phase space region occupied by each of them. Now comes
the first strong assumption: if we suppose that the $N$-particles distribution $f^{(N)}$ is
symmetric in the particles coordinates $\Gamma_1\dots\Gamma_N$, the averaged contribution to the
total force exerted on the test particle is equal for each one, and we have
\begin{eqnarray*}
\displaystyle \sum_{b=2}^N \int f^{(N)} \frac{\partial U_{1b}}{\partial {\bf r}_1}d\Gamma_2\dots
d\Gamma_N = 
\end{eqnarray*}
\begin{equation}
  = (N-1)\int f^{(N)} \frac{\partial U_{12}}{\partial {\bf r}_1} d\Gamma_2\dots d\Gamma_N.
\label{eq:hyp_indist}
\end{equation}

Accordingly, we define the two-particles distribution function as
\begin{equation}
  f^{(2)} = \frac{N!}{(N-2)!}\int f^{(N)}(\Gamma_1,\dots,\Gamma_N)d\Gamma_3\dots d\Gamma_N,
\end{equation}
thus obtaining
\begin{equation}
  \frac{\partial f}{\partial t} + \frac{\partial
    f}{\partial{\bf r}_1} \frac{{\bf p}_1}{m}
  =\int \frac{\partial U_{12}}{\partial {\bf r}_1} \frac{\partial f^{(2)}(t,\Gamma_1,\Gamma_2)}{\partial {\bf p}_1} d\Gamma_2.
\end{equation}
Now comes the second strong assumption, that of molecular chaos, according to which
\begin{equation}
  f^{(2)}(t, \Gamma_1,\Gamma_2) = f(t, \Gamma_1)f(t, \Gamma_2),
\label{eq:hyp_mol_chaos}
\end{equation}
and we finally obtain the Vlasov equation:
\begin{equation}
   \frac{\partial f}{\partial t} + \frac{\partial f}{\partial {\bf r}}{\bf v}
- \frac{\partial \phi}{\partial {\bf r}} \frac{\partial f}{\partial {\bf v}}  = 0,
\end{equation}
where we used ${\bf p} = m{\bf v}$ and the mean potential $\phi({\bf r})$ is given by
\begin{equation}
  \phi({\bf r},t) = \frac{1}{m}\int U(|{\bf r} - {\bf r}'|)f({\bf r}',{\bf p}',t)  d{\bf r}'d{\bf p}'.
\end{equation}

\section*{Acknowledgments}
We thank Viktor Jahnke and Henrique S. Xavier for useful discussions regarding the initial idea of
this paper and S. Salinas for showing us the pioneering works of M. Schönberg. We also thank
Guillaume Plum and Douglas Heggie for discussions about the numerical calculation of the density
profiles. Finally, we thank the referee for the careful reading and helpful suggestions that really
improved the quality of the manuscript. LB is supported by FAPESP. ML and LS are partially
supported by FAPESP and CNPq. LB also thanks the organizers of Gravasco IHP Trimester, during which
part of this work was done, and the kind hospitality of Institut Henri Poincaré in Paris.



\bibliography{/Users/leandro/Dropbox/LeandroSilva/references_leandro}




\end{document}